\pdfoutput=1

\documentclass{pasa}

\usepackage{epsf,psfig}
\usepackage{natbib}

\title[Influence of Dark Matter in Galaxies]{Unveiling the Influence of Dark Matter in Axially Symmetric Galaxies}
\author[Caranicolas and Zotos]{Nicolaos D. Caranicolas \and Euaggelos E. Zotos\thanks{Corresponding author's E-mail: evzotos@physics.auth.gr}\\
~\\
\affil{Department of Physics, Section of Astrophysics, Astronomy \and Mechanics,
Aristotle University of Thessaloniki, \\
GR-541 24, Thessaloniki, Greece \\
~\\
(Received July 2, 2013; Accepted August 4, 2013)}}

\jid{PASA}
\doi{10.1017/pasa.2013.27}
\jyear{\the\year}

\begin{document}

\begin{abstract}
We investigate the regular or chaotic nature of orbits of stars moving in the meridional plane $(R,z)$ of an axially symmetric galactic model with a dense, massive spherical nucleus and a dark matter halo component. In particular, we study the influence of the fractional portion of the dark matter, by computing in each case the percentage of chaotic orbits, as well as the percentages of orbits of the main regular resonant families. In an attempt to distinguish between regular and chaotic motion, we use the Fast Lyapunov Indicator (FLI) method to extensive samples of orbits obtained by integrating numerically the equations of motion as well as the variational equations. Furthermore, a technique which is based mainly on the field of spectral dynamics that utilizes the Fourier transform of the time series of each coordinate is used for identifying the various families of regular orbits and also to recognize the secondary resonances that bifurcate from them. Two cases are studied in our work: (i) the case where we have a disk galaxy model (ii) the case where our model represents an elliptical galaxy. Comparison with early related work is also made.
\end{abstract}


\maketitle

\section{INTRODUCTION}
\label{intro}

From time immemorial, the movements of stars have attracted the interest of the first astronomers. Several decades have been past, since astronomers have turned their interest to the kinematics of stars in galaxies and in galaxy clusters in general. Undoubtedly, galaxies are the basic units of the Universe. An interesting task is to determine the amount of matter in galaxies and, consequently, the amount of matter of the Universe. In particular, there are two key events that are closely associated with this interesting endeavour. The first is the pioneer research of J. Oort who studied the motion of stars in the neighborhood of the Sun and realized that in fact the stars were pulled by forces stronger than those that would be caused only by the visible matter \citep{O24}. On the other hand, the astronomer F. Zwicky discovered that the fast motion of galaxies in clusters cannot be justified only by the visible mass \citep{Z33}. In other words, clusters of galaxies should be scrapped unless the their mass was much greater. The existence of dark matter is the most acceptable scenario to resolve these concerns. Things were clarified when \citet{RF70} presented observational evidence supporting the existence of dark matter in galaxies.

Knowing the rate ratio between luminous and dark matter in galaxies (disk or ellipticals), as well as the relations connecting the fundamental quantities which characterize both components is of great importance, since it allow us to understand how galaxies born and evolve and also how dark matter influences these procedures. Today, it is widely accepted that dark matter is the dominant element in galaxies, taking into account that the vast majority of the total matter of the Universe, 80\% according to today measurements, is indeed dark. Dark matter seems to interacts through gravity. Moreover, apart form gravitational, no other type of dark matter interactions has been observed. This strongly indicates that dark matter interactions should be very weak, probably much more weaker than the particle physics weak interactions. The basic proposed candidates corresponding to dark matter are: (i) neutrinos (Hot Dark Matter), (ii) Warm Dark Matter (WDM), (iii) Cold Dark Matter (CDM) and finally (iv) weak interactive massive particles (WIMPS).

A strong indication for the presence of dark matter in galaxies is derived from their flattened rotation curves at large radii. Using Kepler's third law we have $\Theta(R) = \sqrt{GM(R)/R}$, where $\Theta$ is the rotational velocity at a radius $R$, $G$ is the gravitational constant, while $M(R)$ is the total mass within radius $R$. Conducting observations at large galactocentric distances, where no luminous galactic component was present astronomers found, instead of declining at the expected rate $\Theta(R) \propto R^{-1/2}$, which is true if only $M = const.$, the velocity curves $\Theta(R)$ appear to be flattened. However, $\Theta(R) = const.$, implies that $M(R) \propto R$. This strongly suggests that the mass of galaxies continues to grow, even when there is no luminous component to account for this increase. Moreover, the profiles of the velocity curves indicate that the distribution of light does not match the distribution of mass. In other words, in each galaxy the mass to light ratio $M/L$ increases with the radius $R$. The circular velocity needed for the construction of the rotation curve is obtained, as usually, by measuring the 21 cm emission line from neutral hydrogen HI \citep{RFT80,B81,C85,PS95,HS97}.

The existence of dark matter in elliptical galaxies has been confirmed using observational data derived either from hot gas or their X-ray emission \citep[e.g.,][]{LW99,HBG06,JCOR09,DGCZ10} or even using strong lensing methods \citep[e.g.,][]{RK05,GTR07,KBT09,FAT11}. On the other hand, for the disk galaxies there is a plethora of scenarios describing the formation and the evolution of disk galaxies in correlation with dark matter \citep[e.g.,][]{DSS97,FAH97,AFH98,vdB98,AF00}. A large number of measurements of luminous and dark matter have become available over the last years \citep[e.g.,][]{C06,Cap12,WCT12}. Using these data in combination with dynamically modeling we can reconstruct and therefore study the orbital structure of galaxies.

Today we know that galaxies, contain large amounts of dark matter and, therefore, a further study could provide important information about this invisible matter. A simple way to do this, is to construct models of galaxies containing dark matter. The study of the dynamical properties of these models will provide interesting and useful information, which combined with data from observation will aid significantly in finding a solution for the problem of dark matter. The reader can find interesting information in the field of fitting mass models to the kinematics of disk and elliptical galaxies in a series of papers \citep[see, e.g.][]{BSK04,C07,dB08,GT09}. Another important point, that needs to be emphasized, is the difference between the distribution of dark matter in galaxies and clusters. Observational data \citep[e.g.,][]{S04} suggest that dark matter increases as we move away from the centre to the outer parts of the galaxies. On the contrary in galaxy clusters the dark matter distribution decreases with increasing distance from the galactic centre.

Even today, dark matter is still an open and controversial issue in Astronomy. This is true, because the standard model of cosmology (SMoC) seems to be incompatible with a large amount of data derived from extragalactic observations and modified gravity theories. The reader can find more interesting and detailed information regarding this issue in \citet{K12}.

Taking into account all the above, there is no doubt, that dark matter plays an important role in the dynamical behavior of galaxies. On this basis, it seems of particular interest to build an analytical dynamical model describing the motion of stars both in disk and elliptical galaxies containing dark matter and then study, how the presence and the amount of dark matter affects the regular or chaotic nature of orbits as well as the behavior of the different families of orbits.

The present article is organized as follows: In Section \ref{galmod} we present in detail the structure and the properties of our gravitational galactic model. In Section \ref{compmeth} we describe the computational methods we used in order to determine the character of orbits. In the following Section, we investigate how the parameter corresponding to the fractional portion of the dark matter in galaxies influences the character of the orbits, in both disk and elliptical galaxy models. Our paper ends with Section \ref{disc}, where the discussion and the conclusions of this research are presented.

\section{DESCRIPTION OF THE GALACTIC MODEL}
\label{galmod}

In order to study the dynamical properties of galaxies astronomers often construct galactic models. A galactic model is usually a mathematical expression giving the potential or the mass density of the galaxy, as a function of the distance. The reader can find interesting models, describing motion in galaxies, in \citet{BT08}. Potential density pairs for galaxies were also presented by \citet{VL05}. Over the years, many galactic models have been proposed in order to model the orbital properties in axially symmetric systems. A simple yet realistic axisymmetric logarithmic potential was introduced in \citet{Bin81} for the description of galactic haloes at which the mass density drops like $R^{-2}$ \citep[see also][]{E93}. However, the most well-known model for cold dark matter (CDM) haloes is the flattened cuspy Navarro-Frenk-White (NFW) model \citep{NFW96,NFW97}, where the density at large radii falls like $R^{-1}$ . This model being self-consistent has a major advantage and that's why it is mainly used for conducting $N$-body simulations. Our model, on the other hand, is not self-consistent but simple and contrived, in order to give us the ability to study in detail the orbital behavior of the galactic system. Nevertheless, contrived models can provide an insight into more realistic stellar systems, which unfortunately are very difficult to be studied if we take into account all the astrophysical aspects. Due to the fast that our gravitational model consists of Plummer type potentials, at large galactocentric distances the mass density decreases following the $R^{-5}$ law.

In the present work, we shall investigate how the presence of the dark matter influences the character of the orbits in the meridional plane of an axially symmetric galaxy model with a spherical nucleus and a dark matter halo component. We shall use the usual cylindrical coordinates $(R, \phi, z)$, where $z$ is the axis of symmetry.

The total potential $V(R,z)$ in our model consists of three components: the main galaxy potential $V_{\rm g}$, the central spherical component $V_{\rm n}$ and the dark matter halo component $V_{\rm h}$. The first one is represented by the new mass model introduced in \citet{C12} and is given by
\begin{equation}
V_{\rm g}(R,z) = \frac{-G M_{\rm g}(1 + \delta)}{R_{\rm g}},
\label{Vg}
\end{equation}
where
\begin{equation}
R_{\rm g} = \sqrt{\left(1+\delta\right)\beta^2 + R^2 + \left(\alpha + \sqrt{h^2 + \left(1+\delta^2\right) z^2}\right)^2}.
\label{Rg}
\end{equation}
Here $G$ is the gravitational constant, $M_{\rm g}$ is the mass of the galaxy, $\delta$ is the fractional portion of the dark matter in the galaxy, while $\beta$ represents the core radius of the halo of the disk. The shape of the galaxy is controlled by the parameters $\alpha$ and $h$ which correspond to the horizontal and vertical scale length of the galaxy respectively. Therefore, this potential allow us to describe a variety of galaxy types from a flat disk galaxy when $\alpha, \beta \gg h$ to an oblate elliptical galaxy when $h \gg \alpha, \beta$.

For the description of the spherically symmetric nucleus we use a Plummer potential \citep[e.g.,][]{BT08}
\begin{equation}
V_{\rm n}(R,z) = \frac{-G M_{\rm n}}{\sqrt{R^2 + z^2 + c_{\rm n}^2}},
\label{Vn}
\end{equation}
where $M_{\rm n}$ and $c_{\rm n}$ are the mass and the scale length of the nucleus, respectively. This potential has been used successfully in the past in order to model and therefore interpret the effects of the central mass component in a galaxy \citep[see, e.g.][]{HN90,HPN93,Z12a}. At this point, we must make clear that Eq. (\ref{Vn}) is not intended to represent the potential of a black hole nor that of any other compact object, but just the potential of a dense and massive nucleus thus, we do not include relativistic effects. The dark matter halo is modelled by a similar spherically symmetric potential
\begin{equation}
V_{\rm h}(R,z) = \frac{-G M_{\rm h}}{\sqrt{R^2 + z^2 + c_{\rm h}^2}},
\label{Vh}
\end{equation}
where in this case, $M_{\rm h}$ and $c_{\rm h}$ are the mass and the scale length of the halo, respectively. The spherical shape of the dark halo is simply an assumption, due to the fact that galactic halos may have a variety of shapes.

In this work, we use the well known system of galactic units, where the unit of length is 1 kpc, the unit of mass is $2.325 \times 10^7 {\rm M}_\odot$ and the unit of time is $0.9778 \times 10^8$ yr. The velocity unit is 10 km/s, the unit of angular momentum (per unit mass) is 10 km kpc s$^{-1}$, while $G$ is equal to unity. Finally, the energy unit (per unit mass) is 100 km$^2$s$^{-2}$. In these units, the values of the involved parameters are: $M_{\rm g} = 8200$, $M_n = 400$, $c_{\rm n} = 0.25$, $M_{\rm h} = 7560$ and $c_{\rm h} = 15$. For the disk model we choose: $\beta = 6, \alpha = 3$ and $h = 0.3$, while for the elliptical model we have set $\beta = 2$, $\alpha = 0.5$ and $h = 11.5$. The fractional portion of dark matter $\delta$, on the other hand, is treated as a parameter and its value varies in the interval $0 \leq \delta \leq 0.5$.

\begin{figure*}
\centering
\resizebox{0.9\hsize}{!}{\includegraphics{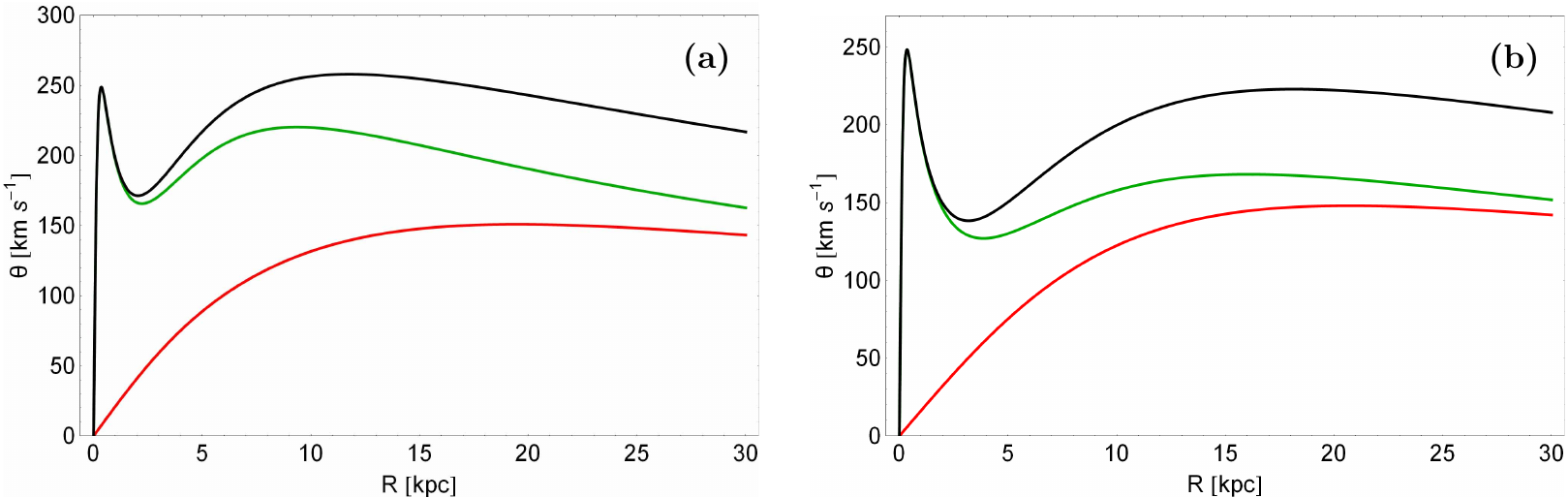}}
\caption{A plot of the rotation curve in our (a-left): disk and (b-right) elliptical galactic model. We can distinguish the total circular velocity (black) and also the contributions from luminous matter (green) and that of the dark matter (red).}
\label{rotvel}
\end{figure*}

\begin{figure*}
\centering
\resizebox{0.9\hsize}{!}{\includegraphics{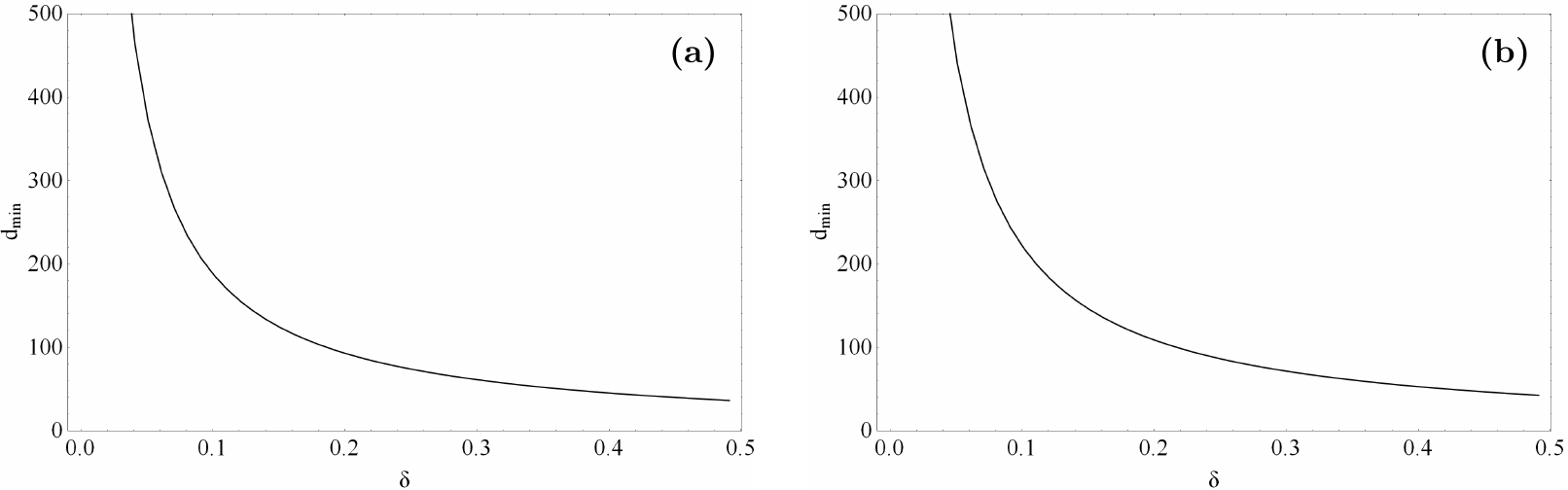}}
\caption{The evolution of the minimum distance $d_{\rm min}$ where negative density appears as a function of the parameter $\delta$ for (a-left): the disk galaxy model (b-right): the elliptical galaxy model.}
\label{deneg}
\end{figure*}

It is well known, that in disk galaxies the circular velocity in the galactic plane $z=0$,
\begin{equation}
\theta(R) = \sqrt{R\left|\frac{\partial V}{\partial R}\right|_{z=0}},
\label{cvel}
\end{equation}
is a very important physical quantity. A plot of $\theta(R)$ for both disk and elliptical galactic model when $\delta = 0.1$ is presented in Fig. \ref{rotvel}(a-b), as a black curve. Moreover, in the same plot the green curve is the contribution from the luminous matter, while the red line corresponds to the contribution form the dark matter. It is seen, that in both cases, the component of the rotational curve generated by the dark matter remains flat for large galactocentric distances, while on the other hand, the luminous matter component continues to decline with increasing distance form the galactic center. We also observe, the characteristic local minimum of the rotation curve due to the massive nucleus, which appears at small values of $R$, when fitting observed data to a galactic model \citep[e.g.,][]{IWTS13,GHBL10}.

At this point, we must clarify, that the mass density in our new galaxy model obtains negative values when the distance from the centre of the galaxy described by the model exceeds a minimum distance $d_{min}$, which strongly depends on the parameter $\delta$. Fig. \ref{deneg}(a-b) shows a plot of $d_{\rm min}$ vs $\delta$ for the both disk and elliptical galaxy models. We see, that even when $\delta = 0.5$ the first indication of negative density occurs only when $d_{\rm min} > 40$ kpc, that is almost at the theoretical boundaries of a real galaxy. Here, we must point out, that our gravitational potential is truncated ar $R_{\rm max} = 30$ kpc for both reasons: (i) otherwise the total mass of the galaxy modeled by this potential would be infinite, which is obviously not physical and (ii) to avoid the existence of negative values of density.

Taking into account that the total potential $V(R,z)$ is axisymmetric, the $z$-component of the angular momentum $L_z$ is conserved. With this restriction, orbits can be described by means of the effective potential
\begin{equation}
V_{\rm eff}(R,z) = V(R,z) + \frac{L_z^2}{2R^2}.
\label{veff}
\end{equation}
The $L_z^2/(2R^2)$ term represents a centrifugal barrier; only orbits with small $L_z$ are allowed to pass near the axis of symmetry. The three-dimensional (3D) motion is thus effectively reduced to a two-dimensional (2D) motion in the meridional plane $(R,z)$, which rotates non-uniformly around the axis of symmetry according to
\begin{equation}
\dot{\phi} = \frac{L_z}{R^2},
\end{equation}
where of course the dot indicates derivative with respect to time.

The equations of motion on the meridional plane are
\begin{eqnarray}
\ddot{R} &=& - \frac{\partial V_{\rm eff}}{\partial R}, \nonumber \\
\ddot{z} &=& - \frac{\partial V_{\rm eff}}{\partial z},
\label{eqmot}
\end{eqnarray}
while the equations describing the evolution of a deviation vector $\delta {\bf{w}} = (\delta R, \delta z, \delta \dot{R}, \delta \dot{z})$ which joins the corresponding phase space points of two initially nearby orbits, needed for the calculation of the standard chaos indicators (the FLI in our case) are given by the variational equations
\begin{eqnarray}
\dot{(\delta R)} &=& \delta \dot{R}, \nonumber \\
\dot{(\delta z)} &=& \delta \dot{z}, \nonumber \\
(\dot{\delta \dot{R}}) &=&
- \frac{\partial^2 V_{\rm eff}}{\partial R^2} \delta R
- \frac{\partial^2 V_{\rm eff}}{\partial R \partial z}\delta z,
\nonumber \\
(\dot{\delta \dot{z}}) &=&
- \frac{\partial^2 V_{\rm eff}}{\partial z \partial R} \delta R
- \frac{\partial^2 V_{\rm eff}}{\partial z^2}\delta z.
\label{vareq}
\end{eqnarray}

Consequently, the corresponding Hamiltonian to the effective potential given in Eq. (\ref{veff}) can be written as
\begin{equation}
H = \frac{1}{2} \left(\dot{R}^2 + \dot{z}^2 \right) + V_{\rm eff}(R,z) = E,
\label{ham}
\end{equation}
where $\dot{R}$ and $\dot{z}$ are momenta per unit mass, conjugate to $R$ and $z$ respectively, while $E$ is the numerical value of the Hamiltonian, which is conserved. Therefore, an orbit is restricted to the area in the meridional plane satisfying $E \geq V_{\rm eff}$.

\section{COMPUTATIONAL METHODS}
\label{compmeth}

In order to obtain the mass profiles of galaxies, we have to construct dynamical models describing the main properties of the galaxies. These models can be generated by deploying two main techniques: (i) using superposition of libraries of orbits \citep[e.g.,][]{G03,TSB04,C06} or (ii) using distributions functions \citep[e.g.,][]{DBVZ96,GJSB98,KSGB00}. In the literature there are also other more specialized dynamical models combining kinematic and photometric data. For instance, axially symmetric Schwarzschild models were used by \citet{B06}, while \citet{H08} used Jeans models in order to fit observational data in the X-ray potential introduced by \citet{HBG06}. Furthermore, axisymmetric Schwarzschild models were also used by \citet{SG10} to fit data derived from the Hubble Space Telescope (HST).

In the recent years, Schwarzschild's superposition method \citep{S79} has been heavily utilized by several researchers \citep[e.g.,][]{RZCMC97,G03,TSB04,VME04,KCEMd05,TSB05} in order to study dark matter distributions in elliptical galaxies therefore, we deem it is necessary to recall and describe briefly in a few words the basic points of this interesting method. Initially, $N$ closed cells define the configuration space, while $K$ orbits extracted from a given mass distribution construct the phase space. Then, integrating numerically the equations of motion, we calculate the amount of time spent by each particular orbit in every cell. Thus, the mass of each cell is directly proportional to the total sum of the stay times of orbits in every cell. Using this technique we manage to compute the unknown weights of the orbits, assuming they are not negative.

In our study, we want to know whether an orbit is regular or chaotic. Several chaos indicators are available in the literature; we chose the FLI method. The FLI \citep{FGL97,LF01} has been proved a very fast, reliable and effective tool, which is defined as
\begin{equation}
\rm FLI(t) = \log \| {\bf{w}(t)} \|, \ \ \ \ t \leq t_{max},
\end{equation}
where ${\bf{w}}(t)$ is a deviation vector. For distinguishing between regular and chaotic motion, we need to compute the FLI for a relatively short time interval of numerical integration $t_{max}$. In particular, we track simultaneously the time-evolution of the orbit itself as well as the deviation vector ${\bf{w}}(t)$ in order to compute the FLI. The variational equations (\ref{vareq}), as usual, are used for the evolution of the deviation vector. The main advantage of the FLI method is that only one deviation vector is required to be computed, while in the case of other dynamical indicators like SALI \citep{SABV04} or GALIs \citep{SBA07} more than one deviation vectors are needed. Therefore, by using the FLI method we need considerable less computation time for integrating and classifying massive sets of initial conditions of orbits.

\begin{figure}
\includegraphics[width=\hsize]{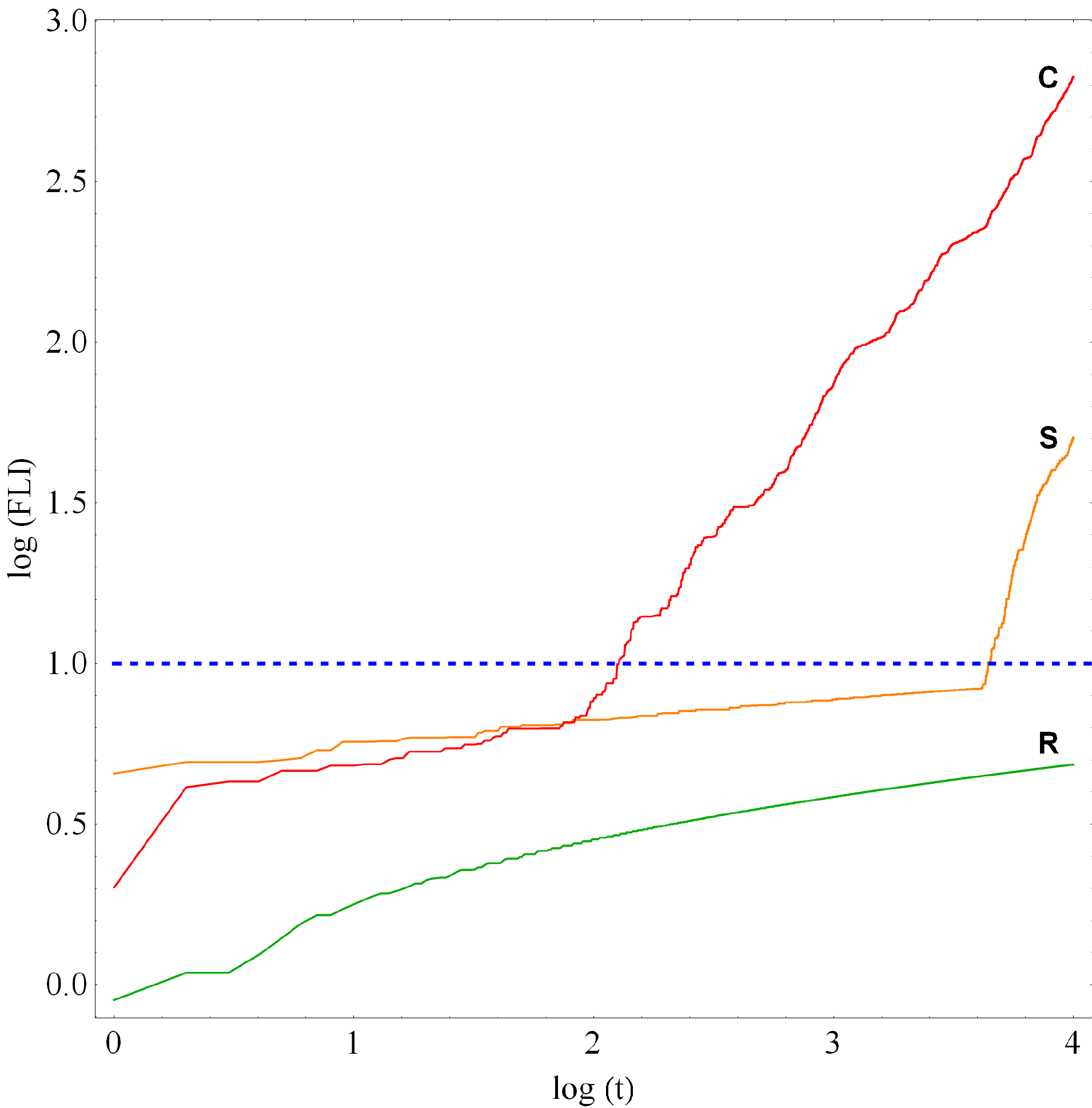}
\caption{Evolution of the FLI of a regular orbit (green color - R), a sticky orbit (orange color - S) and a chaotic orbit (red color - C) in our model for a time period of $10^4$ time units. The horizontal, blue, dashed line corresponds to the threshold value 10 which separates regular from chaotic motion. The chaotic orbits needs only about 130 time units in order to cross the threshold value, while on the other hand, the sticky orbit requires a vast integration time of about 4400 time units so as to reveal its chaotic nature.}
\label{FLIevol}
\end{figure}

The particular time-evolution of the FLI allow us to distinguish fast and safely between regular and chaotic motion as follows: when an orbit is regular the FLI exhibits a linear increase, while on the other hand, in the case of chaotic orbits the FLI evolution is super-exponential. The time evolution of a regular (R) and a chaotic (C) orbit for a time period of $10^4$ time units is presented in Fig. \ref{FLIevol}. We observe, that both regular and chaotic orbits exhibit an increase in the value of FLI but with a complete different rate. Unfortunately, this qualitative criterion is applicable only when someone wants to check the character of individual orbits by plotting and then inspecting by eye the evolution of FLI. Nevertheless, we can easily overcome this drawback by establishing a numerical threshold value, in order to quantify the FLI. After conducting extensive numerical experiments in different types of dynamical systems, we conclude that a safe threshold value for the FLI taking into account the total integration time of $10^4$ time units is the value 10. The horizontal, blue, dashed line in Fig. \ref{FLIevol} corresponds to that threshold value which separates regular from chaotic motion. In order to decide whether an orbit is regular or chaotic, one may use the usual method according to which we check after a certain and predefined time interval of numerical integration, if the value of FLI has become greater than the established threshold value. Therefore, if FLI $\geqslant 10$ the orbit is chaotic, while if FLI $ < 10$ the orbit is regular.

In order to investigate the orbital properties (chaoticity or regularity) of the dynamical system, we need to establish some sample of initial conditions of orbits. The best approach, undoubtedly, would be have been to extract these sample of orbits from the distribution function of the model. Unfortunately, this is not available so, we followed another course of action. For determining the chaoticity of our models, we chose, for each set of values of the parameters of the potential, a dense grid of initial conditions in the $(R,\dot{R})$ phase plane, regularly distributed in the area allowed by the value of the energy $E$. The points of the grid were separated 0.1 units in $R$ and 0.5 units in $\dot{R}$ direction. For each initial condition, we integrated the equations of motion (\ref{eqmot}) as well as the variational equations (\ref{vareq}) with a double precision Bulirsch-Stoer algorithm \citep[e.g.,][]{PTVF92}. In all cases, the energy integral (Eq. (\ref{ham})) was conserved better than one part in $10^{-10}$, although for most orbits, it was better than one part in $10^{-11}$.

Each orbit was integrated numerically for a time interval of $10^4$ time units (10 billion yr), which corresponds to a time span of the order of hundreds of orbital periods but of the order of one Hubble time. The particular choice of the total integration time is an element of paramount importance, especially in the case of the so called ``sticky orbits" (i.e., chaotic orbits that behave as regular ones during long periods of time). A sticky orbit could be easily misclassified as regular by any chaos indicator\footnote{Generally, dynamical methods are broadly split into two types: (i) those based on the evolution of sets of deviation vectors in order to characterize an orbit and (ii) those based on the frequencies of the orbits which extract information about the nature of motion only through the basic orbital elements without the use of deviation vectors.}, if the total integration interval is too small, so that the orbit do not have enough time in order to reveal its true chaotic character. A characteristic example of a sticky orbit (S) in our galactic system can be seen in Fig. \ref{FLIevol}, where we observe that the chaotic character of the particular sticky orbit is revealed after a considerable long integration time of about 4400 time units. Thus, all the set of orbits of the grids of the initial conditions were integrated, as we already said, for $10^4$ time units, thus avoiding sticky orbits with a stickiness at least of the order of a Hubble time. All the sticky orbits which do not show any signs of chaoticity for $10^4$ time units are counted as regular ones, since that vast sticky periods are completely out of scope of our research.

A first step towards the understanding of the overall behavior of our galactic system is knowing the regular or chaotic nature of orbits. Of particular interest, however, is also the distribution of regular orbits into different families. Therefore, once the orbits have been characterized as regular or chaotic, we then further classified the regular orbits into different families, by using the frequency analysis of \citet{CA98,MCW05}. Initially, \citet{BS82,BS84} proposed a technique, dubbed spectral dynamics, for this particular purpose. Later on, this method has been extended and improved by \citet{CA98} and \citet{SN96}. In a recent work, \citet{ZC13} the algorithm was refined even further, so it can be used to classify orbits in the meridional plane. In general terms, this method calculates the Fourier transform of the coordinates of an orbit, identifies its peaks, extracts the corresponding frequencies and search for the fundamental frequencies and their possible resonances. Thus, we can easily identify the various families of regular orbits and also recognize the secondary resonances that bifurcate from them.

Before closing this Section, we would like to make a short note about the nomenclature of orbits. All the orbits of an axisymmetric potential are in fact three-dimensional (3D) loop orbits, i.e., orbits that rotate around the axis of symmetry always in the same direction. However, in dealing with the meridional plane the rotational motion is lost, so the path that the orbit follows onto this plane can take any shape, depending on the nature of the orbit. We will call an orbit according to its behaviour in the meridional plane. Thus, if for example an orbit is a rosette lying in the equatorial plane of the axisymmetric potential, it will be a linear orbit in the meridional plane, etc.

\begin{figure*}
\centering
\resizebox{0.9\hsize}{!}{\includegraphics{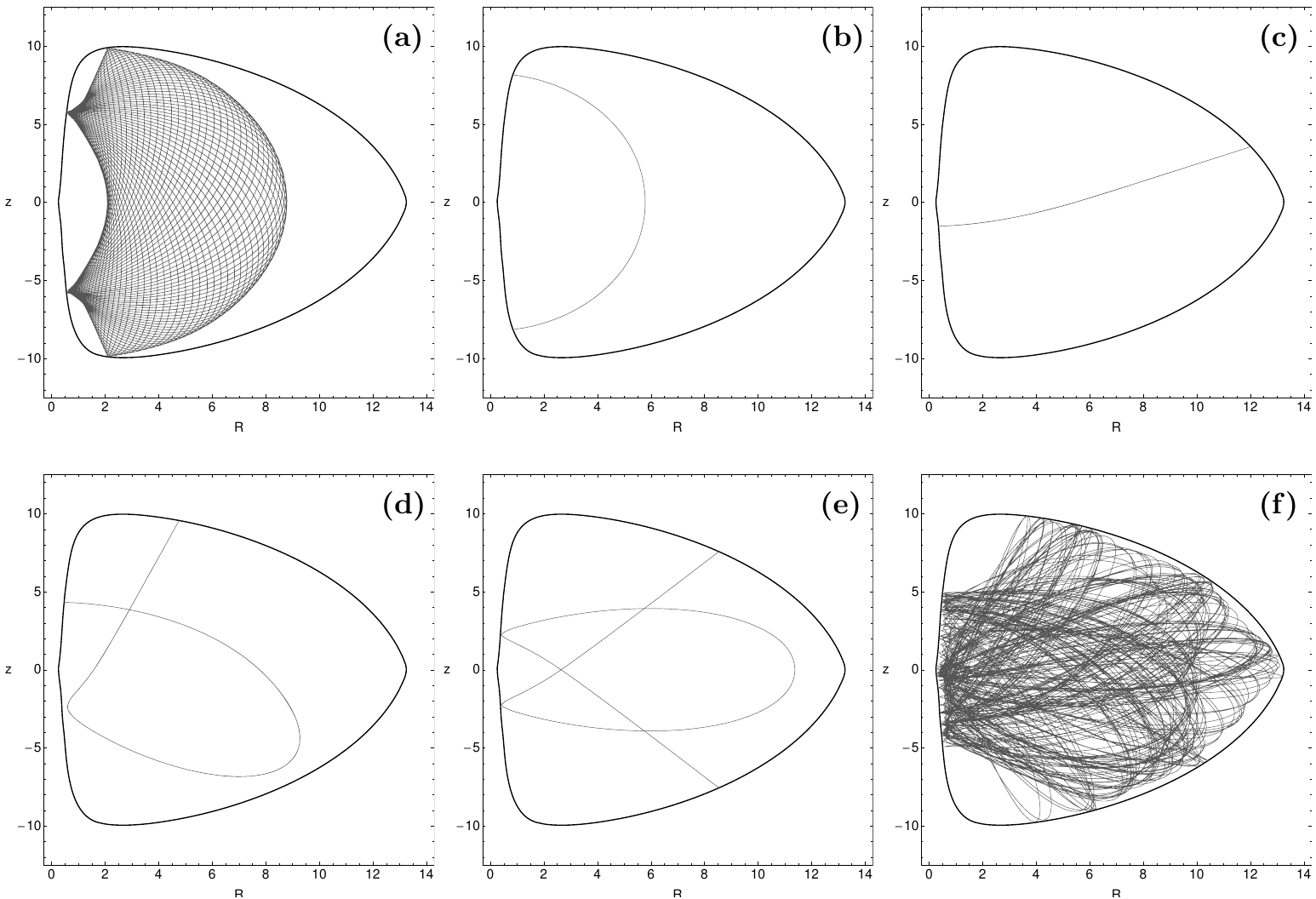}}
\caption{Orbit collection of the six basic types of orbits in the disk galaxy model: (a) box orbit; (b) 2:1 banana-type orbit; (c) 1:1 linear orbit; (d) 3:2 boxlet orbit; (e) 4:3 boxlet orbit; (f) chaotic orbit.}
\label{orbsD}
\end{figure*}

\begin{figure*}
\centering
\resizebox{\hsize}{!}{\includegraphics{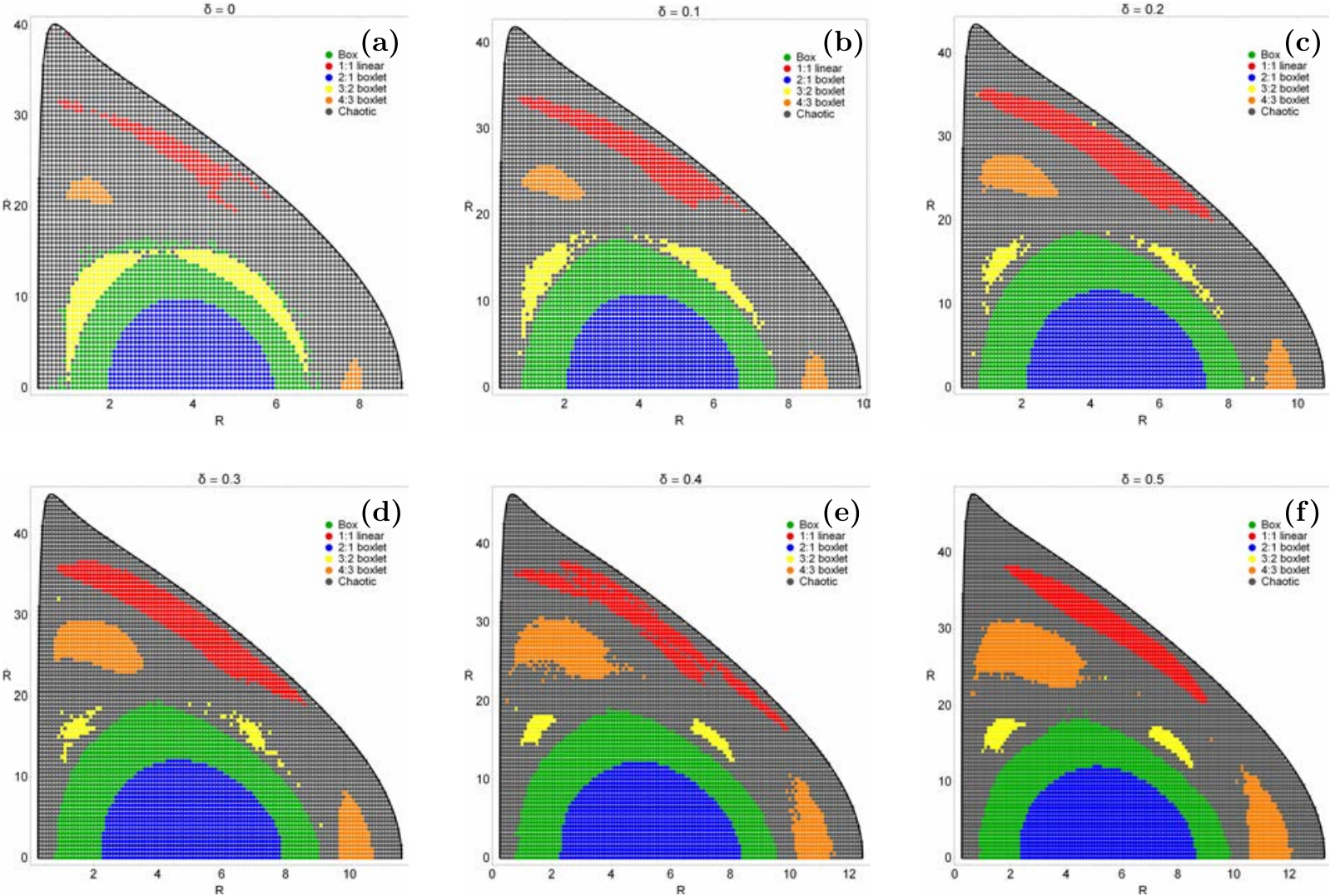}}
\caption{Orbital structure of the $(R,\dot{R})$ phase plane of the disk galaxy model for different values of the fractional portion of the dark matter $\delta$.}
\label{gridsD}
\end{figure*}

\section{ORBIT CLASSIFICATION}
\label{orbclas}

In this Section, we will numerically integrate several sets of orbits in an attempt to distinguish the regular or chaotic nature of motion. We use the initial conditions mentioned in Sec. \ref{compmeth} in order to construct the respective grids of initial conditions, taking always values inside the Zero Velocity Curve (ZVC) defined by
\begin{equation}
\frac{1}{2} \dot{R}^2 + V_{\rm eff}(R,0) = E.
\label{zvc}
\end{equation}
In all cases, the value of the angular momentum of the orbits is $L_z = 15$. We chose for both disk and elliptical galaxy models such energy levels which correspond to $R_{\rm max} \simeq 10$ kpc, where $R_{\rm max}$ is the maximum possible value of $R$ on the $(R,\dot{R})$ phase plane. Once the values of the parameters were chosen, we computed a set of initial conditions as described in Sec. \ref{compmeth} and integrated the corresponding orbits calculating the value of FLI and then classifying the regular orbits into different families.

\subsection{Disk galaxy model}

\begin{table}
\begin{center}
   \caption{Types and initial conditions of the disk galaxy model orbits shown in Figs. \ref{orbsD}(a-f). In all cases, $z_0 = 0$ and $\dot{z_0}$ is found from the energy integral, Eq. (\ref{ham}), while $T_{\rm per}$ is the period of the resonant parent periodic orbits.}
   \label{table1}
   \setlength{\tabcolsep}{3.0pt}
   \begin{tabular}{@{}llccc}
      \hline
      Figure & Type & $R_0$ & $\dot{R_0}$ & $T_{\rm per}$  \\
      \hline
      \ref{orbsD}a & box        &  2.09000000 &  0.00000000 &          - \\
      \ref{orbsD}b & 2:1 banana &  5.79009906 &  0.00000000 & 1.83668227 \\
      \ref{orbsD}c & 1:1 linear &  5.47170604 & 30.97448116 & 1.23526267 \\
      \ref{orbsD}d & 3:2 boxlet &  1.58031642 & 16.62648333 & 3.54124377 \\
      \ref{orbsD}e & 4:3 boxlet & 11.37966251 &  0.00000000 & 4.83447045 \\
      \ref{orbsD}f & chaotic    &  0.40000000 &  0.00000000 &          - \\
      \hline
   \end{tabular}
\end{center}
\end{table}

For the disk galaxy models we choose the energy level $E = -1200$ which is kept constant. Our investigation reveals that in our disk galaxy model there are six main types of orbits: (a) box orbits, (b) 1:1 linear orbits, (c) 2:1 banana-type orbits, (d) 3:2 resonant orbits, (e) 4:3 resonant orbits and (f) chaotic orbits. Note that every resonance $n:m$ is expressed in such a way that $m$ is equal to the total number of islands of invariant curves produced in the $(R,\dot{R})$ phase plane by the corresponding orbit. In Fig. \ref{orbsD}(a-f) we present an example of each of the five basic types of regular orbits, plus an example of a chaotic one. In all cases, we set $\delta = 0.5$. The orbits shown in Figs. \ref{orbsD}a and \ref{orbsD}f were computed until $t = 100$ time units, while all the parent periodic orbits were computed until one period has completed. The black thick curve circumscribing each orbit is the limiting curve in the meridional plane $(R,z)$ defined as $V_{\rm eff}(R,z) = E$. Table \ref{table1} shows the types and the initial conditions for each of the depicted orbits; for the resonant cases, the initial conditions and the period $T_{\rm per}$ correspond to the parent periodic orbits.

\begin{figure}
\includegraphics[width=\hsize]{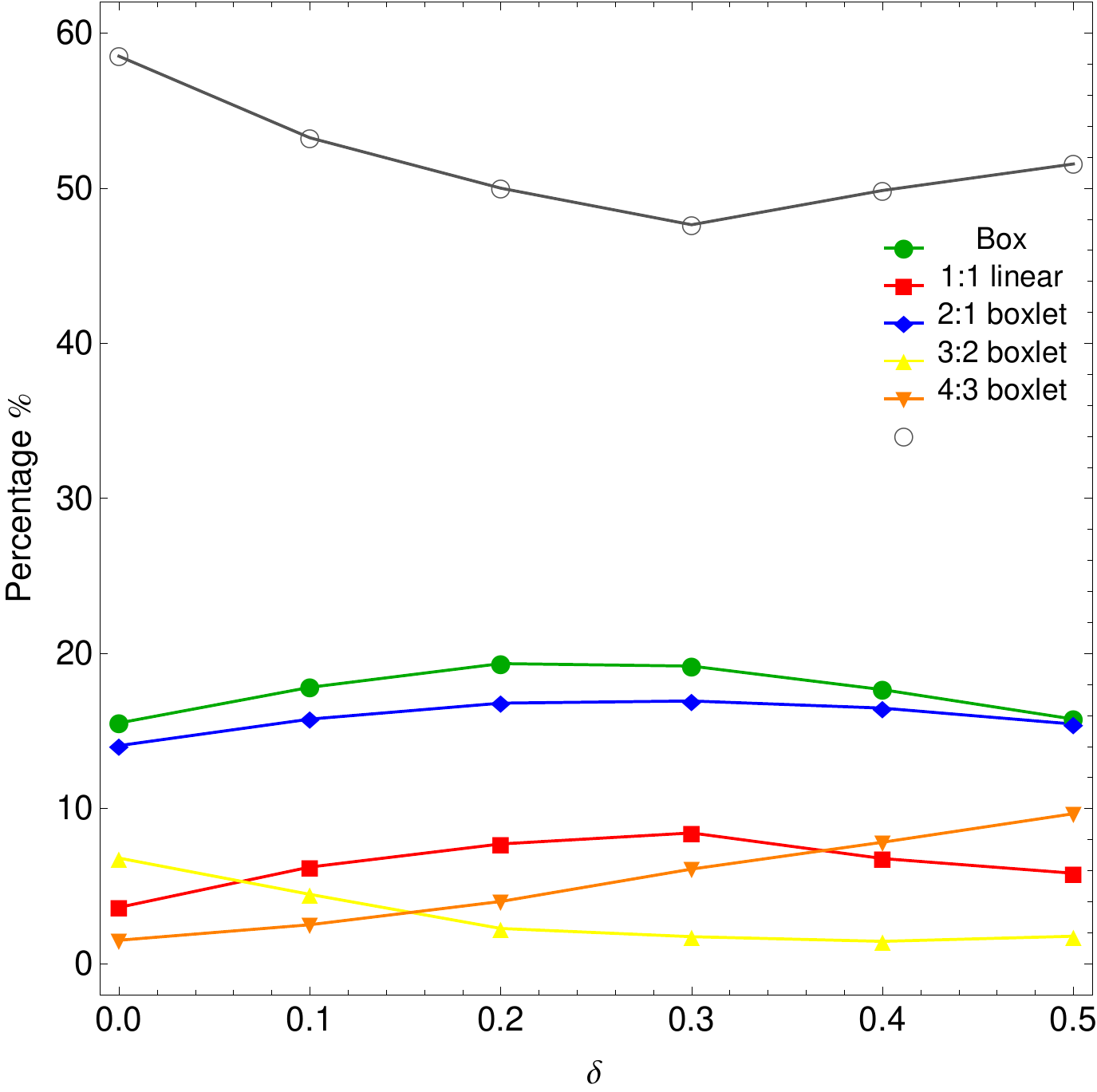}
\caption{Evolution of the percentages of the different kinds of orbits in our disk galaxy model, when varying the fractional portion of the dark matter $\delta$.}
\label{percsD}
\end{figure}

\begin{figure*}
\centering
\resizebox{0.8\hsize}{!}{\includegraphics{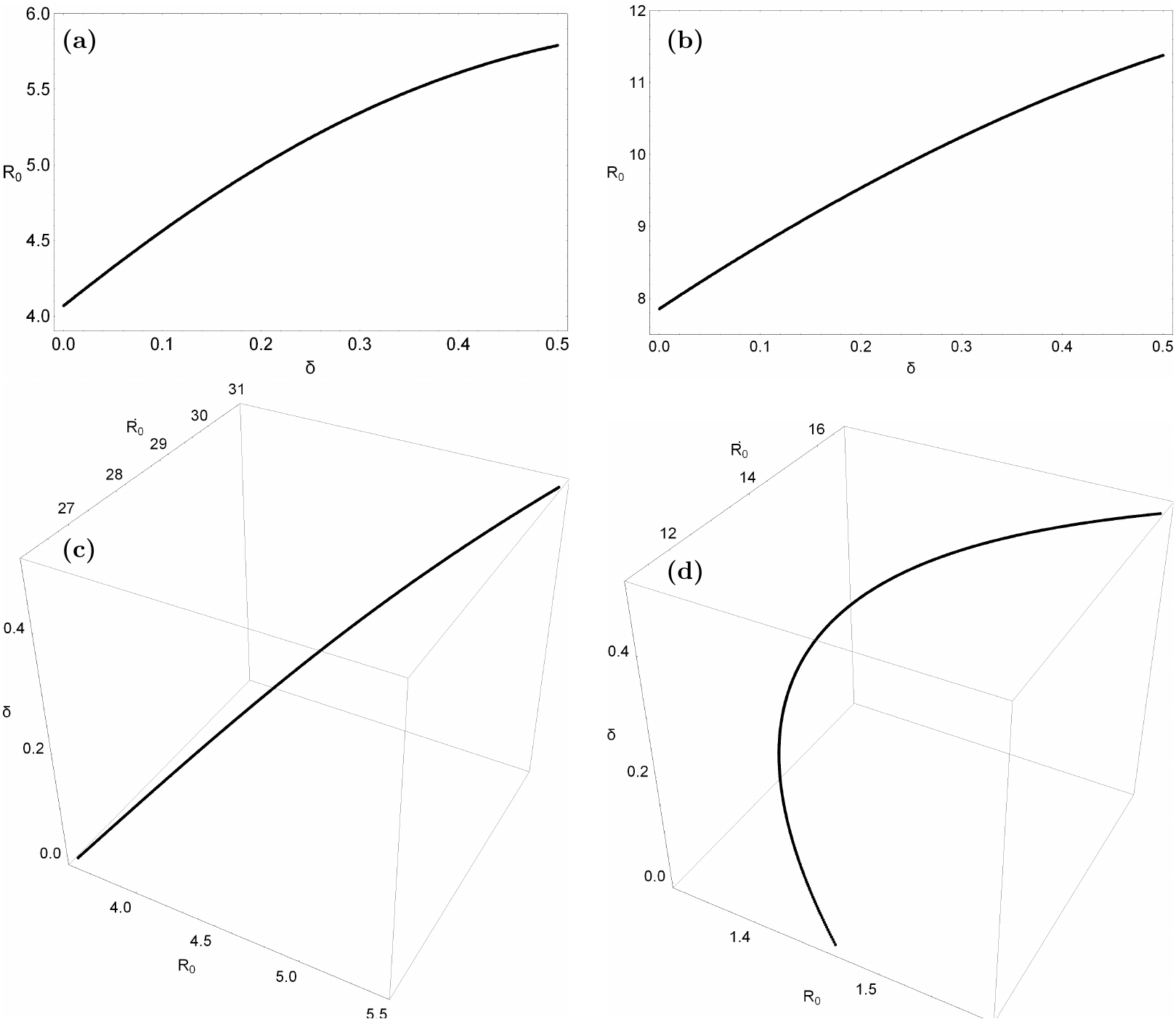}}
\caption{Evolution of the starting position $(R_0,\dot{R_0})$ of the periodic orbits as a function of the fractional portion of the dark matter $\delta$. (a-upper left): 2:1 resonant family; (b-upper right): 4:3 resonant family; (c-lower left): 1:1 resonant family; (d-lower right): 3:2 resonant family.}
\label{FOPD}
\end{figure*}

To study how the fractional portion of the dark matter $\delta$ influences the level of chaos, we let it vary while fixing all the other parameters of our disk galaxy model. As already said, we fixed the values of all the other parameters and integrate orbits in the meridional plane for the set $\delta = \{0,0.1,0.2, ..., 0.5\}$. In all cases, the energy was set to $-1200$ and the angular momentum of the orbits was $L_z=15$. Once the values of the parameters were chosen, we computed a set of initial conditions as described in Sec. \ref{compmeth} and integrated the corresponding orbits computing the FLI of the orbits and then classifying regular orbits into different families.

In Figs. \ref{gridsD}(a-f) we present six grids of orbits that we have classified for different values of the fractional portion of the dark matter $\delta$. Here, we can identify all the different regular families by the corresponding sets of islands which are formed in the phase plane. In particular, we see the five main families already mentioned: (i) 2:1 banana-type orbits surrounding the central periodic point; (ii) box orbits are situated mainly outside of the 2:1 resonant orbits; (iii) 1:1 open linear orbits form the double set of elongated islands in the outer parts of the phase plane; (iv) 3:2 resonant orbits form the double set of islands above the box orbits; and (v) 4:3 resonant orbits correspond to the outer triple set of islands shown in the phase plane. Apart from the regions of regular motion, we observe the presence of a unified chaotic sea which embraces all the islands of stability. The outermost black thick curve is the ZVC defined by Eq. (\ref{zvc}).

Fig. \ref{percsD} shows the resulting percentages of the chaotic orbits and of the main families of regular orbits as $\delta$ varies. It can be seen, that there is a strong correlation between the percentage of most orbits and the value of $\delta$. As the portion of the dark matter increases, there is a gradual decrease in the percentage of chaotic orbit, although this tendency is reversed in models with significant amount of dark matter $(\delta > 0.3)$. In particular, we observe that always chaotic motion is the dominant type of motion and when $\delta > 0.3$ the amount of chaotic orbits grows at the expense of box orbits and 1:1 linear orbits. The meridional 2:1 banana-type orbits, on the other hand, are almost unperturbed by the shifting of the portion of dark matter. Moreover, the 4:3 resonant orbits exhibit a constant increase, while the percentage of the 3:2 resonant orbits remains at very low values. From the diagram shown in Fig. \ref{percsD}, one may conclude that the fractional portion of the dark matter affects mostly the 1:1, 4:3 resonant orbits and chaotic orbits in disk galaxy models.

Of particular interest, is to investigate how the variation in the fractional portion of the dark matter influences the position of the periodic points of the different families of periodic orbits shown in the grids of Fig. \ref{gridsD}. For this purpose, we use the theory of periodic orbits \citep{MH92} and the algorithm developed and applied in \citet{Z13}. In Fig. \ref{FOPD}(a-d) we present the evolution of the starting position of the parent periodic orbits of the four basic families of resonant orbits. The evolution of the 2:1 and 4:3 families shown in Figs. \ref{FOPD}a and Fig. \ref{FOPD}b respectively, is two-dimensional since the starting position $(R_0,0)$ of both families lies on the $R$ axis. On the contrary, studying the evolution of the 1:1 and 3:2 families of periodic orbits is indeed a real challenge due to the peculiar nature of their starting position $(R_0,\dot{R_0})$. In order to visualize the evolution of these families, we need three-dimensional plots such as those presented in Figs. \ref{FOPD}c and \ref{FOPD}d, taking into account the simultaneous relocation of $R_0$ and $\dot{R_0}$. The stability of the periodic orbits can be obtained from the elements of the monodromy matrix $X(t)$ as follows:
\begin{equation}
K = {\rm Tr} \left[X(t)\right] - 2,
\end{equation}
where Tr stands for the trace of the matrix, and $K$ is called the \emph{stability index}. For each set of value of $\delta$, we first located, by means of an iterative process, the position of the parent periodic orbits. Then, using these initial conditions we integrated the variational equations in order to obtain the matrix $X$, with which we computed the index $K$. Our numerical calculations indicate, that in the disk galaxy models, all the different families of periodic orbits remain stable throughout the entire range of the values of $\delta$.

\subsection{Elliptical galaxy model}

\begin{table}
\begin{center}
   \caption{Types and initial conditions of the elliptical galaxy model orbits shown in Figs. \ref{orbsE}(a-h). In all cases, $z_0 = 0$ and $\dot{z_0}$ is found from the energy integral, Eq. (\ref{ham}), while $T_{\rm per}$ is the period of the resonant parent periodic orbits.}
   \label{table2}
   \setlength{\tabcolsep}{3.0pt}
   \begin{tabular}{@{}llccc}
      \hline
      Figure & Type & $R_0$ & $\dot{R_0}$ & $T_{\rm per}$  \\
      \hline
      \ref{orbsE}a & box        &  1.90000000 &  0.00000000 &          - \\
      \ref{orbsE}b & 2:1 banana &  7.18420419 &  0.00000000 &  2.47524161 \\
      \ref{orbsE}c & 1:1 linear &  1.53837422 & 33.45716659 &  1.48665183 \\
      \ref{orbsE}d & 3:2 boxlet &  1.06837848 & 22.64363815 &  4.44068043 \\
      \ref{orbsE}e & 4:3 boxlet & 12.92588226 &  0.00000000 &  5.92689152 \\
      \ref{orbsE}f & 5:3 boxlet &  1.58425772 &  9.55073196 &  7.37479109 \\
      \ref{orbsE}g & 8:5 boxlet & 12.42410864 &  0.00000000 & 11.83080833 \\
      \ref{orbsE}h & chaotic    &  0.40000000 &  0.00000000 &          - \\
      \hline
   \end{tabular}
\end{center}
\end{table}

\begin{figure*}
\centering
\resizebox{0.9\hsize}{!}{\includegraphics{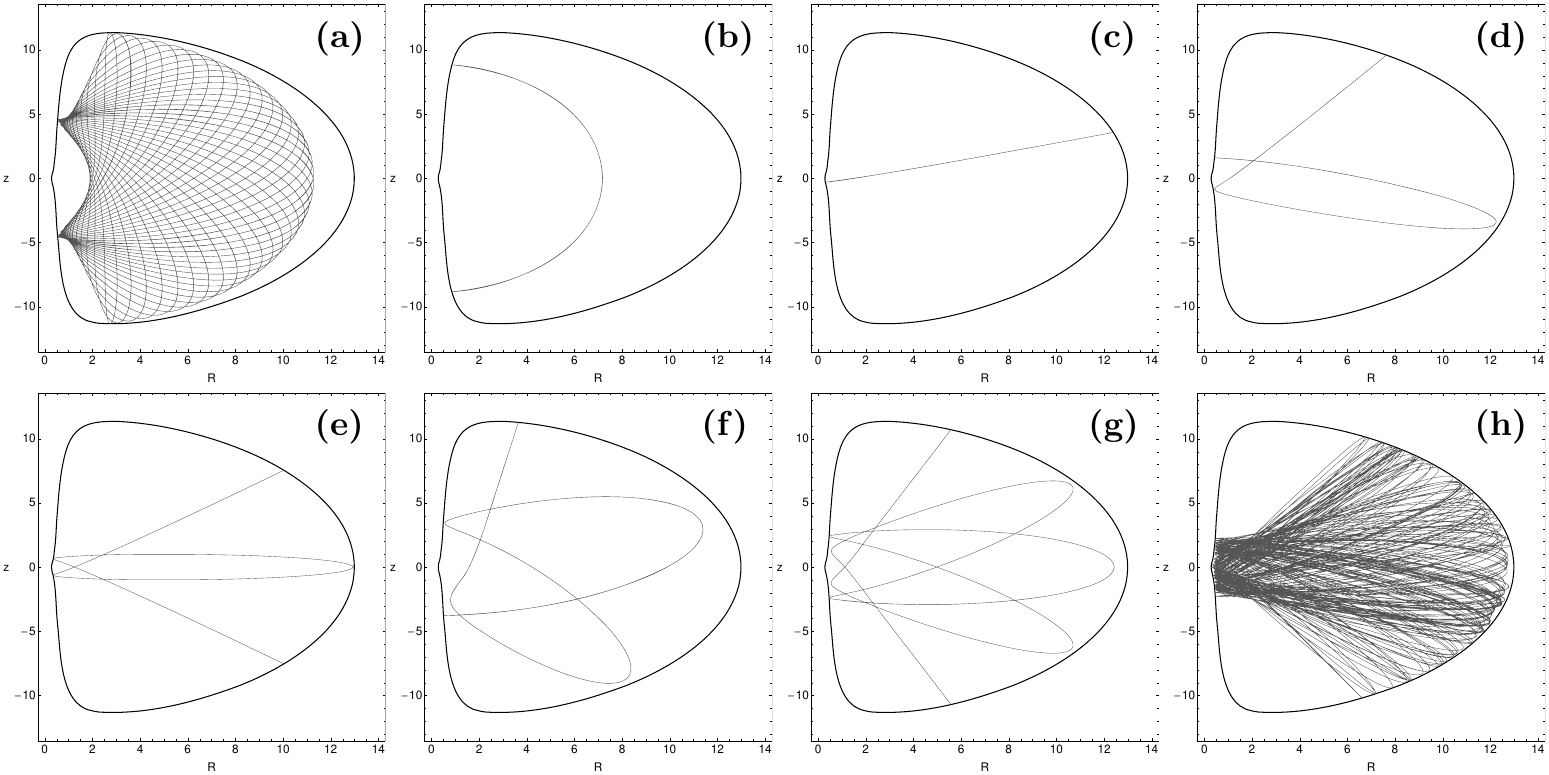}}
\caption{Orbit collection of the eight basic types of orbits in the elliptical galaxy model: (a) box orbit; (b) 2:1 banana-type orbit; (c) 1:1 linear orbit; (d) 3:2 boxlet orbit; (e) 4:3 boxlet orbit; (f) 5:3 boxlet orbit; (g) 8:5 boxlet orbit; (h) chaotic orbit.}
\label{orbsE}
\end{figure*}

In the case of the elliptical galaxy model, we choose the energy level $E = -1100$ which is kept constant. Our numerical investigation shows that in our elliptical galaxy model there are seven main types of orbits: (a) box orbits, (b) 1:1 linear orbits, (c) 2:1 banana-type orbits, (d) 3:2 resonant orbits, (e) 4:3 resonant orbits, (f) 5:3 resonant orbits, (g) 8:5 resonant orbits and (h) chaotic orbits. It is worth noticing, that the basic resonant families, that is the 2:1, 1:1, 3:2 and 4:3 are common in both disk and elliptical galaxy models. However, in the case of the elliptical galaxy additional secondary resonances (i.e. 5:3 and 8:5) appears. Again, every resonance $n:m$ is expressed in such a way that $m$ is equal to the total number of islands of invariant curves produced in the $(R,\dot{R})$ phase plane by the corresponding orbit. In Fig. \ref{orbsE}(a-h) we present an example of each of the seven basic types of regular orbits, plus an example of a chaotic one. In all cases, we set $\delta = 0.5$. The orbits shown in Figs. \ref{orbsE}a and \ref{orbsE}f were computed until $t = 100$ time units, while all the parent periodic orbits were computed until one period has completed. Table \ref{table2} shows the types and the initial conditions for each of the depicted orbits; for the resonant cases, the initial conditions and the period $T_{\rm per}$ correspond to the parent periodic orbits.

\begin{figure*}
\centering
\resizebox{\hsize}{!}{\includegraphics{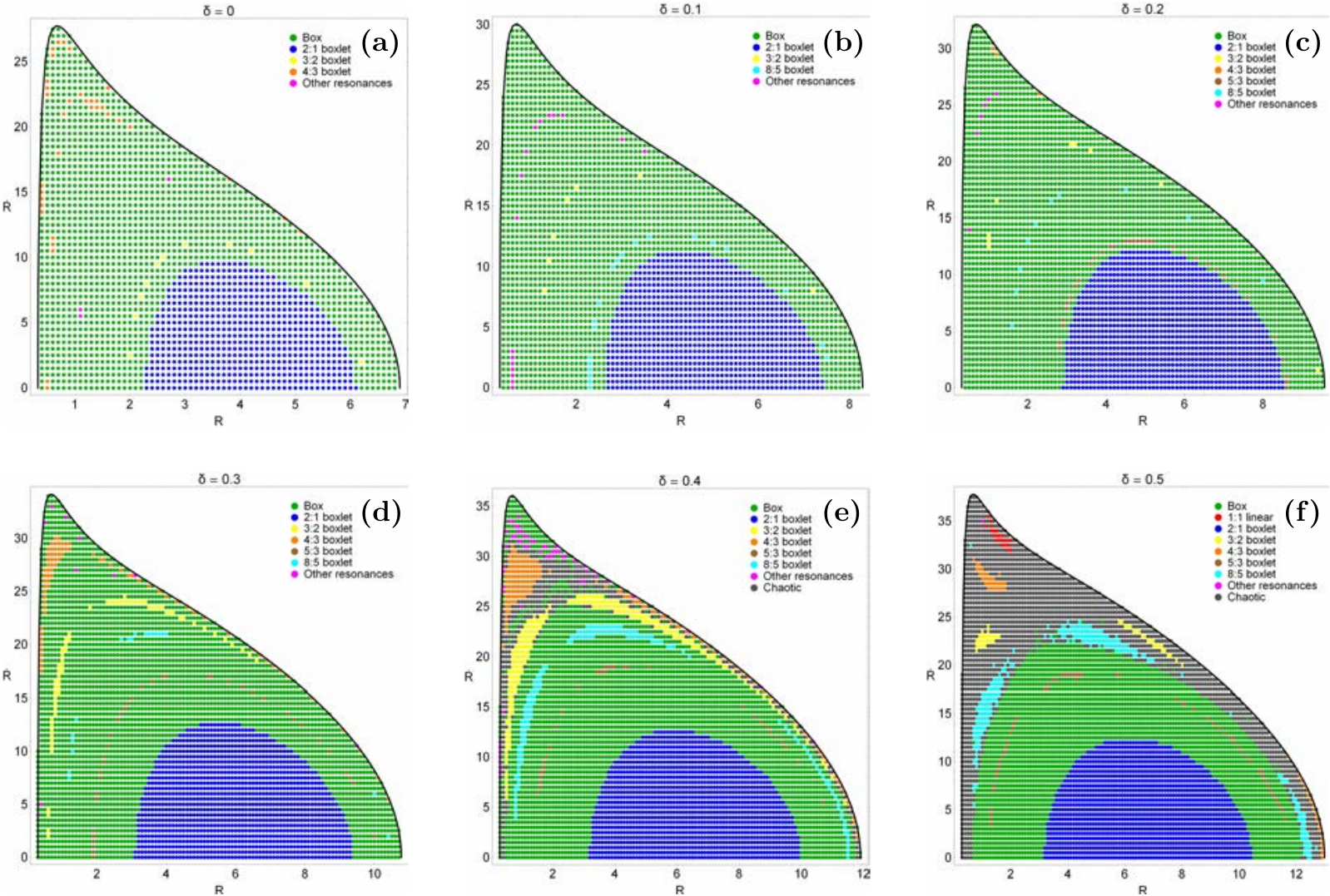}}
\caption{Orbital structure of the $(R,\dot{R})$ phase plane of the elliptical galaxy model for different values of the fractional portion of the dark matter $\delta$.}
\label{gridsE}
\end{figure*}

In order to study how the fractional portion of the dark matter $\delta$ influences the level of chaos, we let it vary while fixing all the other parameters in our elliptical galaxy model. As already said, we fixed the values of all the other parameters and integrate orbits in the meridional plane for the set $\delta = \{0,0.1,0.2, ..., 0.5\}$. In all cases the energy was set to $-1100$ and the angular momentum of the orbits was $L_z = 15$. Once the values of the parameters were chosen, we computed a set of initial conditions as described in Sec. \ref{compmeth} and integrated the corresponding orbits computing the FLI of the orbits and then classifying regular orbits into different families.

Six grids of initial conditions $(R_0,\dot{R_0})$ that we have classified for different values of the fractional portion of the dark matter $\delta$ are shown in Figs. \ref{gridsE}(a-f). By inspecting these grids, we can identify all the different regular families by the corresponding sets of islands which are produced in the phase plane. In particular, we see the seven main families of regular orbits already mentioned: (i) 2:1 banana-type orbits correspond to the central periodic point; (ii) box orbits situated mainly outside of the 2:1 resonant orbits; (iii) 1:1 open linear orbits form the double set of elongated islands in the outer parts of the phase plane; (iv) 3:2 resonant orbits form the double set of islands; (v) 4:3 resonant orbits correspond to the outer triple set of islands shown in the phase plane; (vi) 5:3 resonant orbits forming the set of the three small islands inside the region of box orbits and (vii) 8:5 resonant orbits producing the set of five islands. It is evident, that the structure of the phase plane in the elliptical galaxy models differs greatly from that of the disk models. We observe, that when the amount of dark matter in the elliptical galaxy is low, almost the entire phase plane is covered by different types of regular orbits. On the other hand, chaotic motion appears, mainly at the outer parts of the phase plane, only when the galaxy possesses a significant amount of dark matter ($\delta \geq 0.4$).

\begin{figure}
\includegraphics[width=\hsize]{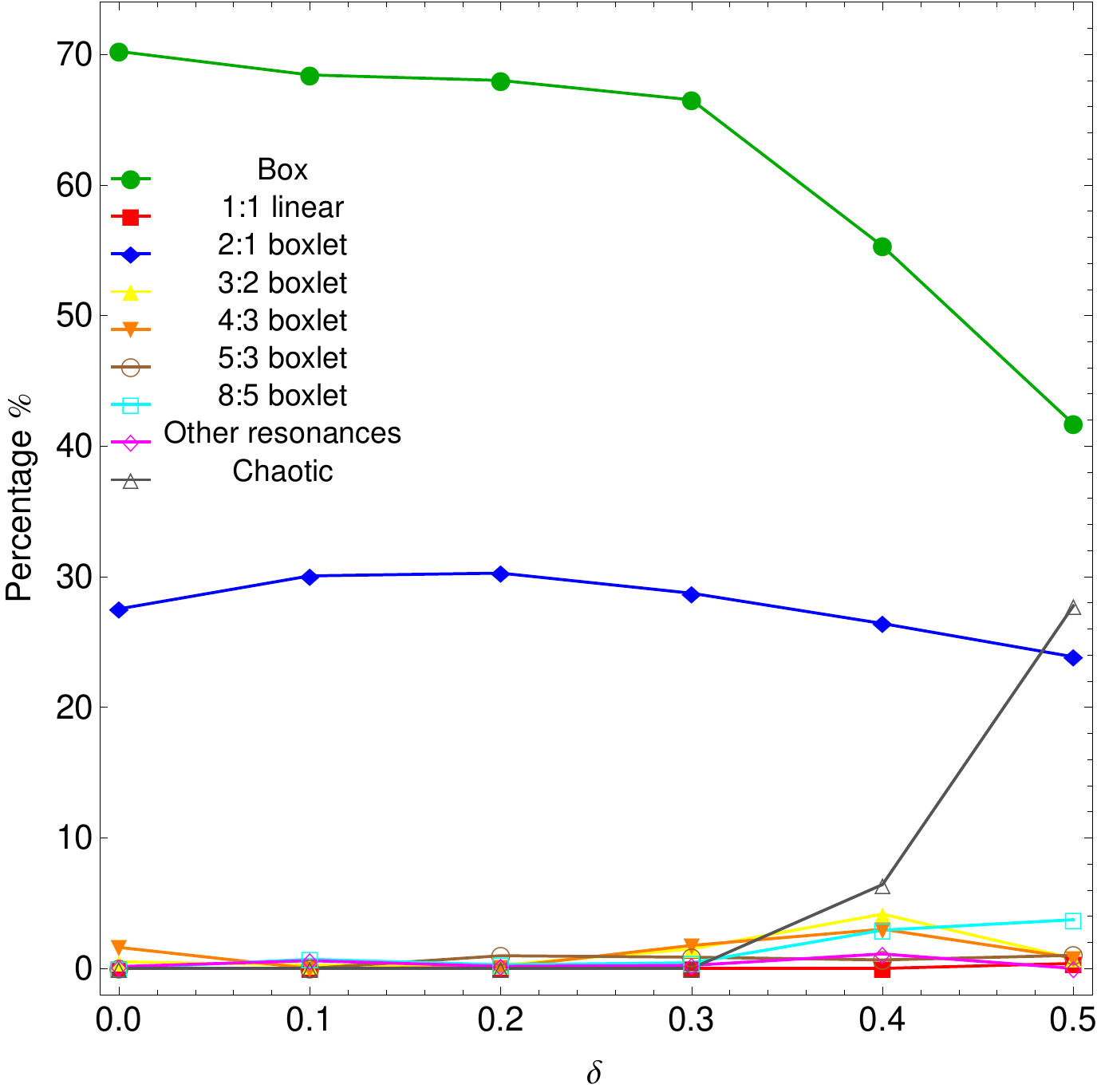}
\caption{Evolution of the percentages of the different kinds of orbits in our elliptical galaxy model, when varying the fractional portion of the dark matter $\delta$.}
\label{percsE}
\end{figure}

In the following Fig. \ref{percsE} we present the resulting percentages of the chaotic orbits and of also the main families of regular orbits as $\delta$ varies. It can be seen, that the motion of stars in elliptical galaxies, is almost entirely regular, being the box orbits the all-dominant type. The percentage of box orbits is however reduced as the portion of the dark matter is increased, although they always remain the most populated family. It is also seen, that the percentages of the 2:1 banana-type orbits exhibits a minor decrease with increasing $\delta$. On the other hand, the chaotic orbits start to grow rapidly as soon as the galaxy contains significant amount of dark matter ($\delta > 0.3$). Moreover, all the other resonant families of orbits are immune to changes of the portion of the dark matter, since their percentages remain almost unperturbed and at very low values (less than 5\%). From Fig. \ref{percsE} one may conclude that dark matter in elliptical galaxies mostly affects box and chaotic orbits. In fact, a portion of box orbits turns into chaotic as the galaxy gains more dark matter.

\begin{figure*}
\centering
\resizebox{0.9\hsize}{!}{\includegraphics{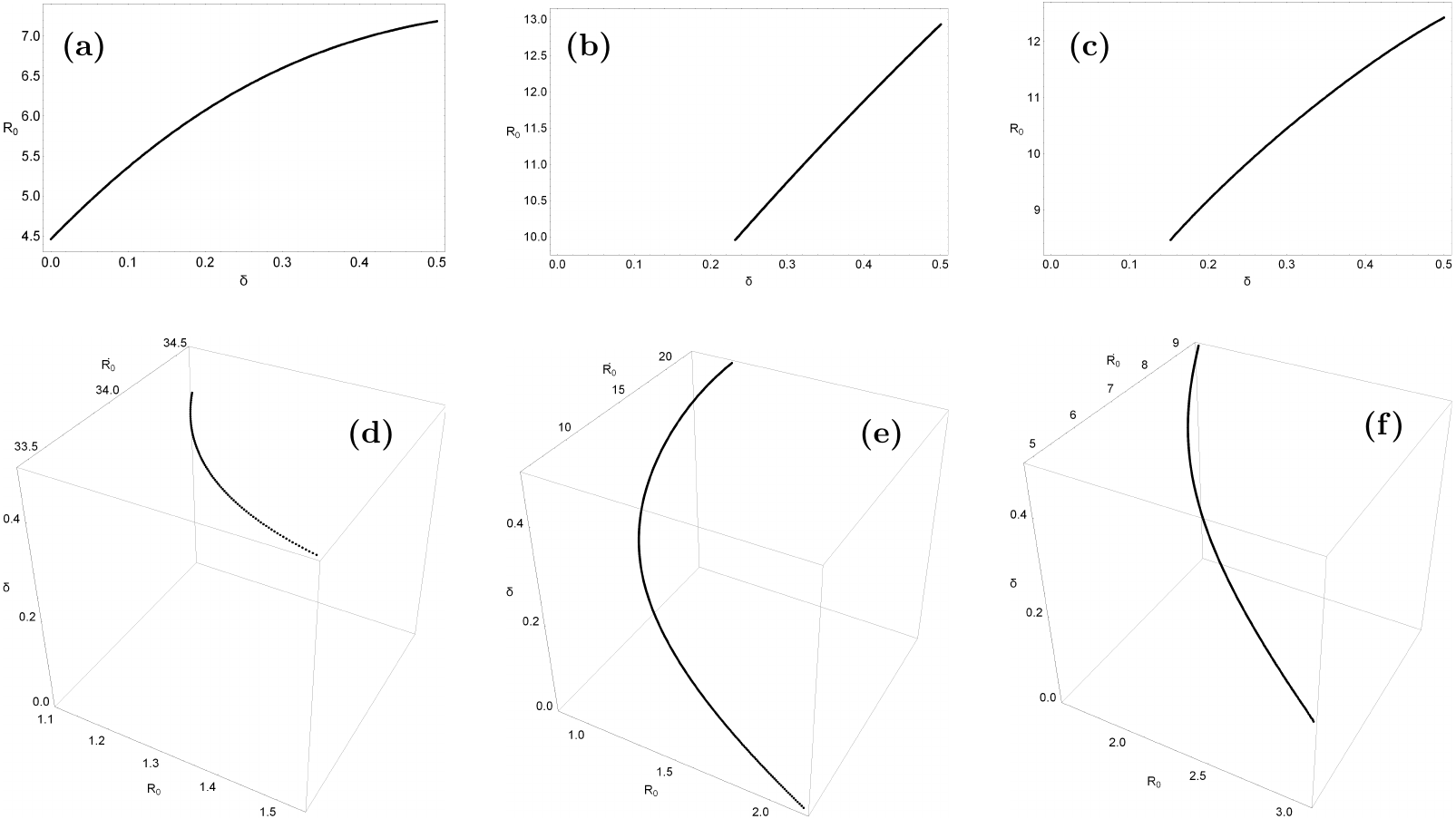}}
\caption{Evolution of the starting position $(R_0,\dot{R_0})$ of the periodic orbits as a function of the fractional portion of the dark matter $\delta$. (a) 2:1 resonant family; (b) 4:3 resonant family; (c) 8:5 resonant family; (d) 1:1 resonant family; (e) 3:2 resonant family; (f) 5:3 resonant family.}
\label{FOPE}
\end{figure*}

We close this Section, by presenting how the variation in the fractional portion of the dark matter influences the position of the periodic points of the different families of periodic orbits shown in the grids of Fig. \ref{gridsE}(a-f). We follow the same method used previously in the case of the disk galaxy. In Fig. \ref{FOPE}(a-f) the evolution of the starting position of the parent periodic orbits of the six basic families of resonant orbits is given. Once more, the evolution of the 2:1, 4:3 and 8:5 families shown in Figs. \ref{FOPE}(a-c), is two-dimensional since the starting position $(R_0,0)$ of these families lies on the $R$ axis. On the contrary, the evolution of the 1:1, 3:2 and 5:3 families are shown in the three-dimensional plots in Figs. \ref{FOPE}(d-f) thus following simultaneous the relocation of $R_0$ and $\dot{R_0}$ as the fractional portion of the dark matter $\delta$ varies. Our numerical calculations suggest, that all the computed resonant periodic orbits were found to be stable. Furthermore, we should point out that several families of periodic orbits in the case of the elliptical galaxy do not cover the entire range of the values of $\delta$.

\section{DISCUSSION AND CONCLUSIONS}
\label{disc}

There is no doubt, that the determination of the nature of dark matter is one of the most interesting and challenging open problems that scientists try to solve. In the present paper, we used the analytic, axisymmetric, mass model which was introduced in \citet{C12} and embraces the general features of a dense, massive nucleus and a spherical dark matter halo. We made this choice because observations show that the assumption of a spherical halo seems to be reasonable. On the other hand, non spherical and triaxial haloes are also possible in some galaxies \citep[see, e.g.][]{GSKVK04,TdWBK08,VC11}. A galaxy with a dark matter halo is undoubtedly a very complex entity and, therefore, we need to assume some necessary simplifications and assumptions in order to be able to study mathematically the orbital behavior of such a complicated stellar system. For this purpose, our model is simple and contrived, in order to give us the ability to study different aspects of the dynamical model. Nevertheless, contrived models can provide an insight into more realistic stellar systems, which unfortunately are very difficult to be studied, if we take into account all the astrophysical aspects. On the other hand, self-consistent models are mainly used when conducting $N$-body simulations. However, this is entirely out of the scope of the present paper. Once again, we have to point out that the simplicity of our model is necessary; otherwise it would be extremely difficult, or even impossible, to apply the extensive and detailed dynamical study presented in this study. Similar gravitational models with the same limitations and assumptions were used successfully several times in the past in order to investigate the orbital structure in much more complicated galactic systems \citep[see, e.g.][]{Z12b,Z13a}.

In this work, we investigated how influential is the parameter corresponding to the portion of the dark matter $\delta$ on the level of chaos and also on the distribution of regular families among its orbits in both disk and elliptical galaxy models. The main results of our research can be summarized as follows:

\textbf{(1).} In disk galaxy models, the fractional portion of the dark matter affects mostly the 1:1, 4:3 resonant orbits and the chaotic orbits, while the effect on all the other resonant families is very weak compared to them. In particular, chaotic motion is always the prevailing type of motion but when the amount of dark matter is high enough, the amount of chaotic orbits grows at the expense of box orbits and 1:1 linear orbits.

\textbf{(2).} That portion of dark matter in elliptical galaxy models influences mainly the box and the chaotic orbits. In fact, box orbits are the dominant family when the amount of dark matter is low, but the percentage of chaotic orbits quickly grows as the dark matter is being accumulated, by collapsing the percentage of box orbits, although they always remain the most populated family.

\textbf{(3).} The percentage of the observed chaos in disk galaxy models with dark matter is significantly larger compared to that in elliptical galaxy models. This result coincide with similar conclusions obtained using different types of dynamical models in order to model disk and elliptical galaxies \citep[see, e.g.][]{PC05,Z11}.

The outcomes of the present research are considered as a promising first step in the task of exploring the orbital structure in both disk and elliptical galaxy models containing dark matter. Taking into account that our results are encouraging, it is in our future plans to modify properly our dynamical model in order to expand our study in three dimensions.

\section*{Acknowledgments}

We would like to express our warmest thanks to the anonymous referee for the careful reading of the manuscript and for all the aptly suggestions and comments which allowed us to improve both the quality and the clarity of our paper.

\end{document}